\documentclass{Rinton-P9x6}

\newcommand{\postscript}[2]{\setlength{\epsfxsize}{#2\hsize}
   \centerline{\epsfbox{#1}}}
\newcommand{\mweak}{M_{\text{weak}}}
\newcommand{\mplanck}{M_{\text{Pl}}}

\newcommand{\kev}{\text{keV}}
\newcommand{\mev}{\text{MeV}}
\newcommand{\gev}{\text{GeV}}
\newcommand{\tev}{\text{TeV}}

\newcommand{\K}{\text{K}}
\newcommand{\s}{\text{s}}

\newcommand{\etal}{{\em et al.}}

\newcommand{\eqref}[1]{Eq.~(\ref{#1})}

\newcommand{\WIMP}{\text{WIMP}}
\newcommand{\SWIMP}{\text{SWIMP}}
\newcommand{\mWIMP}{m_{\WIMP}}
\newcommand{\mSWIMP}{m_{\SWIMP}}
\newcommand{\YWIMP}{Y_{\WIMP}}

\newcommand{\Gravitino}{\tilde{G}}
\newcommand{\Bino}{\tilde{B}}

\newcommand{\stau}{\tilde{\tau}}
\newcommand{\slepton}{\tilde{\ell}}
\newcommand{\epsEM}{\varepsilon_{\text{EM}}}
\newcommand{\epshad}{\varepsilon_{\text{had}}}
\newcommand{\zetaEM}{\zeta_{\text{EM}}}
\newcommand{\zetahad}{\zeta_{\text{had}}}

\usepackage{bm}
\def\agt{\mathrel{\raise.3ex\hbox{$>$\kern-.75em\lower1ex\hbox{$\sim$}}}}
\def\alt{\mathrel{\raise.3ex\hbox{$<$\kern-.75em\lower1ex\hbox{$\sim$}}}}
\newcommand{\text}[1]{{\rm #1}}
\newcommand{\mtext}[1]{\mbox{{\rm #1}}}

\begin{document}

% UCI-TR-2003-40

\title{SuperWIMPs in Supergravity
%\footnote{Invited talk presented at the International Conference on 20
%Years of SUGRA and the Search for SUSY and Unification (SUGRA20),
%March 2003, Northeastern University, Boston, USA.}  
}

\author{Jonathan L.~Feng}

\address{Department of Physics and Astronomy \\
University of California, Irvine, CA 92697, USA
}

\maketitle

\abstracts{ In supergravity theories, a natural possibility is that
neutralinos or sleptons freeze out at their thermal relic density, but
then decay to gravitinos after about a year.  The resulting gravitinos
are then superWIMPs --- superweakly-interacting massive particles that
naturally inherit the desired relic density from late decays of
conventional WIMPs.  SuperWIMP dark matter escapes all conventional
searches.  However, the late decays that produce superWIMPs provide
new and promising early universe signatures for cold dark matter.  }

\section{Introduction}
\label{sec:introduction}

One of the many virtues of supergravity theories, such as minimal
supergravity,\cite{Chamseddine:jx} is that they naturally provide an
excellent cold dark matter candidate --- the
neutralino.\cite{Goldberg:1983nd} Neutralinos emerge as the lightest
supersymmetric particle (LSP) in simple models.  Assuming $R$-parity
conservation, they are stable.  In addition, their thermal relic
density is naturally in the range required for dark matter, and there
are promising prospects for both direct and indirect detection.

Recently we have explored an alternative solution to the dark matter
problem in supergravity theories.\cite{Feng:2003xh,Feng:2003uy}
Neutralino dark matter is a possibility only if the gravitino mass
$m_{\tilde{G}}$ is above the neutralino mass.  In supergravity, this
need not be the case --- $m_{\tilde{G}}$ and the scalar and gaugino
masses are all of the order of $\mweak \sim \langle F \rangle /
\mplanck$, where $\mweak \sim 100~\gev - 1~\tev$, $\mplanck \simeq 1.2
\times 10^{19}~\gev$, and $\langle F \rangle$ are the weak, Planck,
and supersymmetry-breaking scales, respectively.  The specific
ordering depends on unknown, presumably ${\cal O}(1)$, coefficients.

The gravitino is therefore the LSP in roughly ``half'' of parameter
space, and it may be cold dark matter.  Assuming that the universe
inflates and then reheats to a temperature below $\sim 10^8~\gev -
10^{10}~\gev$,\cite{Krauss:1983ik} the number of gravitinos is
negligible after reheating.  Then, because the gravitino couples only
gravitationally with all interactions suppressed by $\mplanck$, it
plays no role in the thermodynamics of the early universe.  The
next-to-lightest supersymmetric particle (NLSP) therefore freezes out
as usual; if it is weakly-interacting, its relic density will be near
the desired value.  However, much later, after
\begin{equation}
\tau \sim \frac{\mplanck^2}{\mweak^3} \sim 10^5~\s - 10^{8}~\s \ ,
\label{year}
\end{equation}
the WIMP decays to the LSP, converting much of its energy density to
gravitinos.  Gravitino LSPs may therefore form a significant relic
component of our universe, with a relic abundance naturally near
$\Omega_{\text{DM}} \simeq 0.23$.\cite{Spergel:2003cb} Models with
weak-scale extra dimensions also provide a similar dark matter
particle in the form of Kaluza-Klein
gravitons,\cite{Feng:2003xh,Feng:2003nr} with Kaluza-Klein gauge
bosons or leptons playing the role of the decaying WIMP.\cite{KKDM}
Because such dark matter candidates naturally preserve the WIMP relic
abundance, but have interactions that are weaker than weak, we refer
the whole class of such particles as ``superWIMPs.''

The superWIMP possibility differs markedly from previous proposals for
gravitino dark matter.  In the original gravitino dark matter
scenario, gravitinos have thermal equilibrium abundances and form warm
dark matter.\cite{Pagels:ke} The required $\sim \kev$ mass for such
gravitinos is taken as evidence for a new intermediate supersymmetry
breaking scale $\langle F \rangle$.  Alternatively, weak-scale
gravitinos may be produced with the correct abundances during
reheating, provided that the reheat temperature is tuned
appropriately.  In contrast to these scenarios, the properties of
superWIMP dark matter are determined by the two known mass scales
$\mweak$ and $\mplanck$.  SuperWIMP dark matter therefore preserves
the main quantitative virtue of conventional WIMPs, naturally
connecting the electroweak scale to the observed relic density. In
addition, the mechanism of gravitino production through late decays
implies that the superWIMP scenario is highly predictive, and, as we
shall see, testable.

\section{SuperWIMP Properties}
\label{sec:superwimp}

As outlined above, superWIMP dark matter is produced in decays $\WIMP
\to \SWIMP + S$, where $S$ denotes one or more standard model
particles.  The superWIMP is essentially invisible, and so the
observable consequences rely on finding signals of $S$ production in
the early universe.  In principle, the strength of these signals
depend on what $S$ is and its initial energy distribution.  For the
parameters of greatest interest here, however, $S$ quickly initiates
electromagnetic or hadronic cascades.  As a result, the observable
consequences depend only on the WIMP's lifetime $\tau$ and the average
total electromagnetic or hadronic energy released in WIMP
decay.\cite{Ellis:1984er,Ellis:1990nb,Kawasaki:1994sc,%
Holtmann:1998gd,Kawasaki:2000qr,Asaka:1998ju,Cyburt:2002uv,BBNhad}

In many simple supergravity models, the lightest standard model
superpartner is a Bino-like neutralino.  For pure Binos,
\begin{equation}
\Gamma(\tilde{B} \to \gamma \Gravitino) 
= \frac{\cos^2\theta_W}{48\pi M_*^2}
\frac{m_{\tilde{B}}^5}{m_{\Gravitino}^2} 
\left[1 - \frac{m_{\Gravitino}^2}{m_{\tilde{B}}^2} \right]^3 
\left[1 + 3 \frac{m_{\Gravitino}^2}{m_{\tilde{B}}^2} \right] \ .
\label{Binolifetime}
\end{equation}
This decay width, and all results that follow, includes the
contributions from couplings to both the spin $\pm 3/2$ and $\pm 1/2$
gravitino polarizations.  These must all be included, as they are
comparable in models with high-scale supersymmetry breaking.  In the
limit $\Delta m \equiv \mWIMP - \mSWIMP \ll \mSWIMP$,
$\Gamma(\tilde{B} \to \gamma \Gravitino) \propto (\Delta m)^3$ and the
decay lifetime is
\begin{equation}
\tau(\Bino \to \gamma \Gravitino) 
\approx 2.3 \times 10^7~\s 
\left[ \frac{100~\gev}{\Delta m} \right]^3  \ ,
\end{equation}
independent of the overall $\mWIMP$, $\mSWIMP$ mass scale.

If a slepton is the lightest standard model superpartner, its decay
width is
\begin{equation}
 \Gamma(\slepton \to \ell \Gravitino)
 =\frac{1}{48\pi M_*^2} 
\frac{m_{\slepton}^5}{m_{\Gravitino}^2} 
\left[1 - \frac{m_{\Gravitino}^2}{m_{\slepton}^2} \right]^4 .
\label{sleptonlifetime}
\end{equation}
This expression is valid for any scalar superpartner decaying to a
nearly massless standard model partner.  In particular, it holds for
$\slepton = \tilde{e}$, $\tilde{\mu}$, or $\tilde{\tau}$, and
arbitrary mixtures of the $\slepton_L$ and $\slepton_R$ gauge
eigenstates.  In the limit $\Delta m \equiv \mWIMP - \mSWIMP \ll
\mSWIMP$, the decay lifetime is
\begin{equation}
\tau(\slepton \to \ell \Gravitino) 
 \approx 3.6 \times 10^8~\s 
\left[ \frac{100~\gev}{\Delta m} \right]^4 
\frac{m_{\Gravitino}}{1~\tev} \ . 
\end{equation}

The electromagnetic energy release is conveniently written in terms of
\begin{equation}
\zetaEM \equiv \epsEM \YWIMP \ ,
\end{equation}
where $\epsEM$ is the initial electromagnetic energy released in each
WIMP decay, and $\YWIMP \equiv n_{\WIMP}/n_{\gamma}^{\text{BG}}$ is
the number density of WIMPs before they decay, normalized to the
number density of background photons $n_{\gamma}^{\text{BG}} = 2
\zeta(3) T^3/\pi^2$.  We define hadronic energy release similarly as
$\zetahad \equiv \epshad \YWIMP$. In the superWIMP scenario, WIMP
velocities are negligible when they decay.  We will be concerned
mainly with the case where $S$ is a single nearly massless particle,
and so we define
\begin{equation}
E_S \equiv \frac{\mWIMP^2 - \mSWIMP^2}{2\mWIMP} \ ,
\label{ES}
\end{equation}
the potentially visible energy in such cases.  

For the neutralino WIMP case, $S = \gamma$. (Possible contributions to
hadronic decay products are discussed in the last section.)  Clearly
all of the initial photon energy is deposited in an electromagnetic
shower, and so
\begin{equation}
\epsEM = E_{\gamma} \ , \quad \epshad \simeq 0 \ .
\label{Egamma}
\end{equation}
For the slepton case, the energy release is flavor-dependent.  For the
case of staus,
\begin{equation}
\epsEM \approx 
\frac{1}{3} E_{\tau} - E_{\tau} \ , \quad \epshad = 0 \ ,
\label{Etau}
\end{equation}
where the range in $\epsEM$ results from the possible variation in
electromagnetic energy from $\pi^{\pm}$ and $\nu$ decay products.  The
precise value of $\epsEM$ is in principle calculable once the stau's
chirality and mass, and the superWIMP mass, are specified.  However,
as the possible variation in $\epsEM$ is not great relative to other
effects, we will simply present results below for the representative
value of $\epsEM = \frac{1}{2} E_{\tau}$.

\begin{figure}[tbp]
\postscript{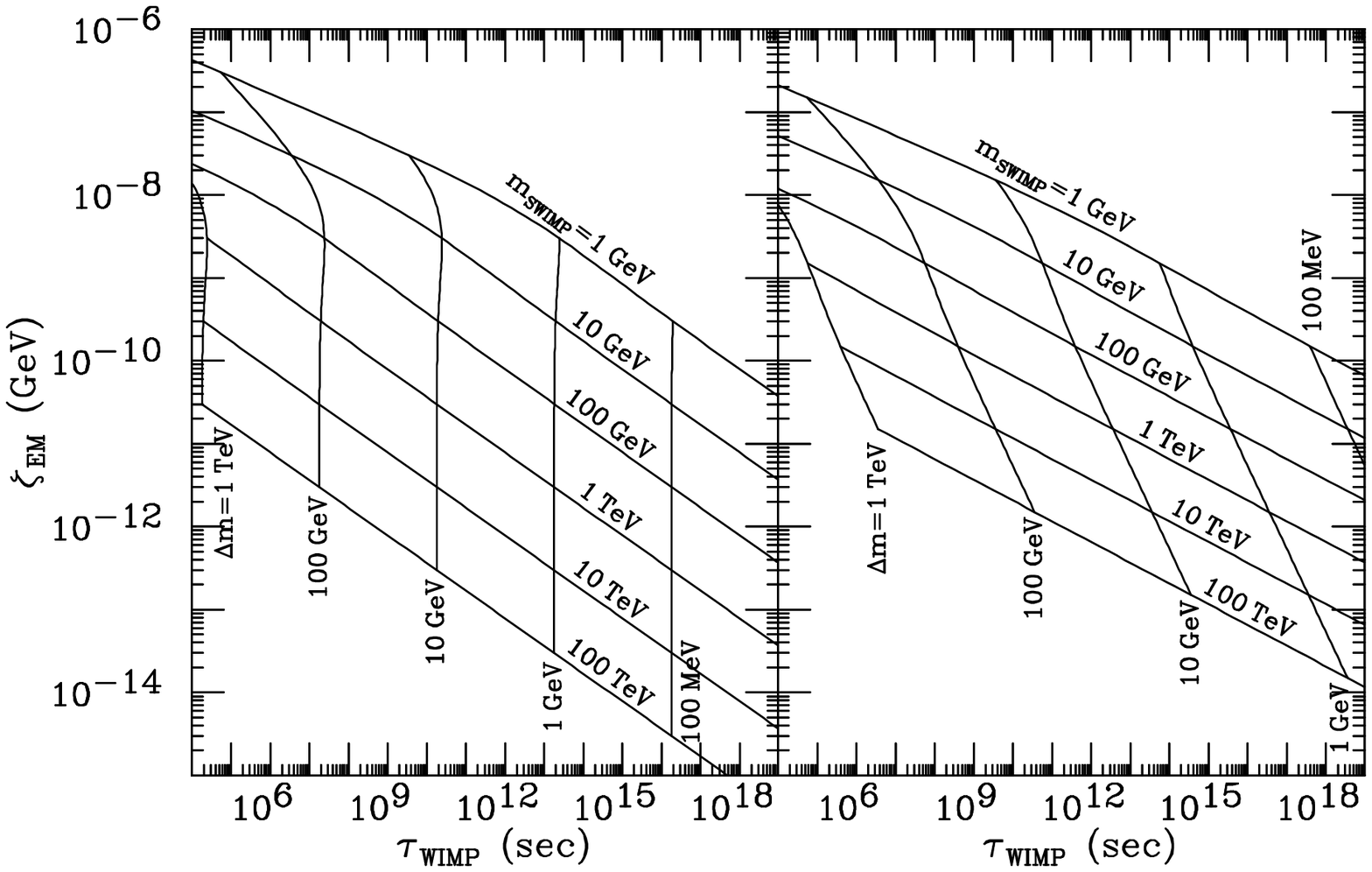}{0.85}
\caption{Predicted values of WIMP lifetime $\tau$ and electromagnetic
  energy release $\zetaEM \equiv \epsEM \YWIMP$ in the $\Bino$ (left)
  and $\stau$ (right) WIMP scenarios for $\mSWIMP = 1~\gev$,
  $10~\gev$, \ldots, $100~\tev$ (top to bottom) and $\Delta m \equiv
  \mWIMP - \mSWIMP = 1~\tev$, $100~\gev$, \ldots, $100~\mev$ (left to
  right).  For the $\stau$ WIMP scenario, we assume $\epsEM =
  \frac{1}{2} E_{\tau}$.
\label{fig:prediction} }
\end{figure}

The lifetimes and energy releases in the Bino and stau WIMP scenarios
are given in Fig.~\ref{fig:prediction} for a range of $(\mWIMP, \Delta
m)$.  For natural weak-scale values of these parameters, the lifetimes
and energy releases in the neutralino and stau scenarios are similar,
with lifetimes of about a year, in accord with the rough estimate of
\eqref{year}, and energy releases of
\begin{equation}
\zeta_{\text{EM}} \sim 10^{-9}~\gev \ .
\label{zeta}
\end{equation}
Such values have testable implications, as we discuss in the following
two sections.

\section{Big Bang Nucleosynthesis}
\label{sec:bbn}

\subsection{Standard BBN and CMB Baryometry}

Big Bang nucleosynthesis predicts primordial light element abundances
in terms of one free parameter, the baryon-to-photon ratio $\eta
\equiv n_B / n_{\gamma}$.  At present, the observed D, $^4$He, $^3$He,
and $^7$Li abundances may be accommodated for baryon-to-photon ratios
in the range\cite{Hagiwara:fs}
\begin{equation}
\eta_{10} \equiv  \eta / 10^{-10} = 2.6-6.2 \ .
\label{etarange}
\end{equation}
In light of the difficulty of making precise theoretical predictions
and reducing (or even estimating) systematic uncertainties in the
observations, this consistency is a well-known triumph of standard Big
Bang cosmology.

At the same time, given recent and expected advances in precision
cosmology, the standard BBN picture merits close scrutiny. Recently,
BBN baryometry has been supplemented by CMB data, which alone yields
$\eta_{10} = 6.1 \pm 0.4$.\cite{Spergel:2003cb}  Observations of
deuterium absorption features in spectra from high redshift quasars
imply a primordial D fraction of $\text{D/H} = 2.78_{-0.38}^{+0.44}
\times 10^{-5}$.\cite{Kirkman:2003uv}  Combined with standard BBN
calculations,\cite{Burles:2000zk} this yields $\eta_{10} = 5.9 \pm
0.5$.  The remarkable agreement between CMB and D baryometers has two
new implications for scenarios with late-decaying particles.  First,
assuming there is no fine-tuned cancellation of unrelated effects, it
prohibits significant entropy production between the times of BBN and
decoupling.  Second, the CMB
measurement supports determinations of $\eta$ from D, already
considered by many to be the most reliable BBN baryometer.  It
suggests that if D and another BBN baryometer disagree, the
``problem'' lies with the other light element abundance --- either its
systematic uncertainties have been underestimated, or its value is
modified by new astrophysics or particle physics. Such disagreements
may therefore provide specific evidence for late-decaying particles in
general, and superWIMP dark matter in particular.  We address this
possibility here.

In standard BBN, the baryon-to-photon ratio $\eta_{10} = 6.0\pm 0.5$
favored by D and CMB observations predicts\cite{Burles:2000zk}
\begin{eqnarray}
Y_p &=& 0.2478 \pm 0.0010 \label{4He} \\
^3\text{He/H} &=& (1.03 \pm 0.06) \times 10^{-5} \label{3He} \\
^7\text{Li/H} &=& 4.7_{-0.8}^{+0.9} \times 10^{-10} \label{Li}
\end{eqnarray}
at 95\% CL, where $Y_p$ is the $^4$He mass fraction. At present the
greatest discrepancy lies in $^7$Li, where all measurements are below
the prediction of \eqref{Li}.  The $^7$Li fraction may be determined
precisely in very low metallicity stars.  Three independent
studies\cite{Thorburn,Bonafacio,Ryan:1999vr} find
\begin{eqnarray}
\mtext{$^7$Li/H} &=& 1.5_{-0.5}^{+0.9} \times 10^{-10} \quad 
\mtext{(95\% CL)} \\
\mtext{$^7$Li/H} &=& 1.72_{-0.22}^{+0.28} \times 10^{-10} \ 
\mtext{($1\sigma + \text{sys}$)} \\
\mtext{$^7$Li/H} &=& 1.23_{-0.32}^{+0.68} \times 10^{-10} \ 
\mtext{(stat + sys, 95\% CL)} \ ,
\end{eqnarray}
where depletion effects have been estimated and included in the last
value.  Within the published uncertainties, the observations are
consistent with each other but inconsistent with \eqref{Li}, with
central values lower than predicted by a factor of $3-4$.  $^7$Li may
be depleted from its primordial value by astrophysical effects, for
example, by rotational mixing in stars that brings Lithium to the core
where it may be burned,\cite{Pinsonneault:1998nf,Vauclair:1998it} but
it is controversial whether this effect is large enough to reconcile
observations with the BBN prediction.\cite{Ryan:1999vr}

The other light element abundances are in better agreement.  For
example, a global analysis,\cite{Burles:2000zk} using the ``high''
$Y_p$ values of Izotov and Thuan,\cite{Izotov} finds $\chi^2 = 23.2$
for 3 degrees of freedom, where $\chi^2$ is completely dominated by
the $^7$Li discrepancy.

\subsection{SuperWIMPs and the $^7$Li Underabundance}

\begin{figure}[tbp]
\postscript{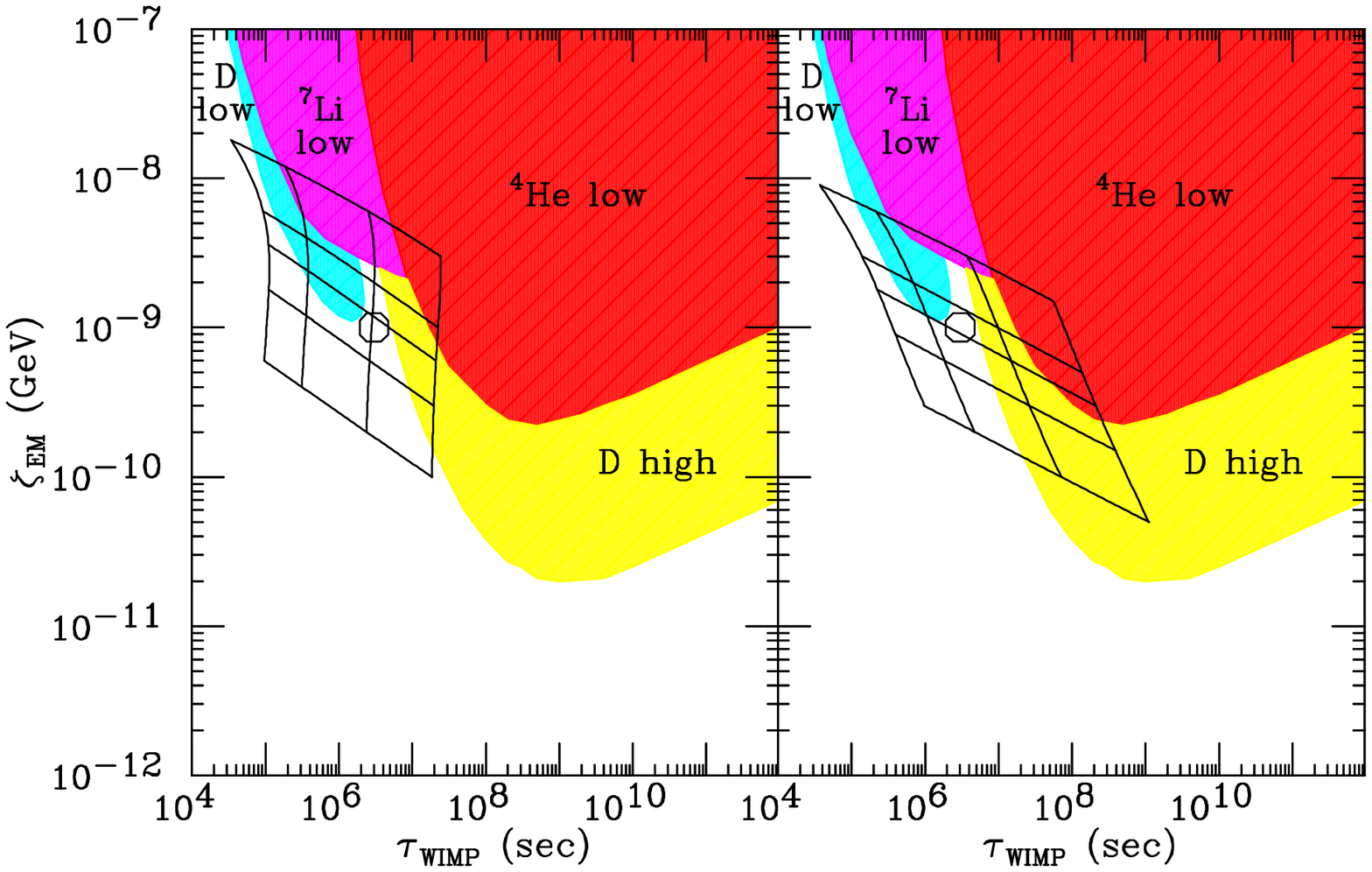}{0.85}
\caption{The grid gives predicted values of WIMP lifetime $\tau$ and
electromagnetic energy release $\zetaEM \equiv \epsEM \YWIMP$ in the
$\Bino$ (left) and $\stau$ (right) WIMP scenarios for $\mSWIMP =
100~\gev$, $300~\gev$, $500~\gev$, $1~\tev$, and $3~\tev$ (top to
bottom) and $\Delta m \equiv \mWIMP - \mSWIMP = 600~\gev$, $400~\gev$,
$200~\gev$, and $100~\gev$ (left to right).  For the $\stau$ WIMP
scenario, we assume $\epsEM = \frac{1}{2} E_{\tau}$. The analysis of
BBN constraints by Cyburt, Ellis, Fields, and
Olive\protect\cite{Cyburt:2002uv} excludes the shaded regions.  The
best fit region with $(\tau, \zetaEM) \sim (3 \times 10^6~\s,
10^{-9}~\gev)$, where $^7$Li is reduced to observed levels by late
decays of WIMPs to superWIMPs, is given by the circle.
\label{fig:bbn} }
\end{figure}

Given the overall success of BBN, the first implication for new
physics is that it should not drastically alter any of the light
element abundances.  This requirement restricts the amount of energy
released at various times in the history of the universe. A recent
analysis by Cyburt, Ellis, Fields, and Olive of electromagnetic
cascades finds that the shaded regions of Fig.~\ref{fig:bbn} are
excluded by such considerations.\cite{Cyburt:2002uv}  The various
regions are disfavored by the following conservative criteria:
\begin{eqnarray}
\mtext{D low}      \ : && \text{D/H} < 1.3 \times 10^{-5} \\
\mtext{D high}     \ : && \text{D/H} > 5.3 \times 10^{-5} \\
\mtext{$^4$He low} \ : && Y_p < 0.227 \\
\mtext{$^7$Li low} \ : && \mtext{$^7$Li/H} < 0.9 \times 10^{-10} \ .
\end{eqnarray}

A subset of superWIMP predictions from Fig.~\ref{fig:prediction} is
superimposed on this plot.  The subset is for weak-scale $\mSWIMP$ and
$\Delta m$, the most natural values, given the independent motivations
for new physics at the weak scale.  The BBN constraint eliminates some
of the region predicted by the superWIMP scenario, but regions with
$\mWIMP, \mSWIMP \sim \mweak$ remain viable.

The $^7$Li anomaly discussed above may be taken as evidence for new
physics, however.  To improve the agreement of observations and BBN
predictions, it is necessary to destroy $^7$Li without harming the
concordance between CMB and other BBN determinations of $\eta$.  This
may be accomplished for $(\tau, \zetaEM) \sim (3 \times 10^6~\s,
10^{-9}~\gev)$.\cite{Cyburt:2002uv}  This ``best fit'' point is
marked in Fig.~\ref{fig:bbn}.  The amount of energy release is
determined by the requirement that $^7$Li be reduced to observed
levels without being completely destroyed -- one cannot therefore be
too far from the ``$^7$Li low'' region.  In addition, one cannot
destroy or create too much of the other elements.  $^4$He, with a
binding threshold energy of 19.8 MeV, much higher than Lithium's 2.5
MeV, is not significantly destroyed.  On the other hand, D is loosely
bound, with a binding energy of 2.2 MeV.  The two primary reactions
are D destruction through $\gamma \text{D} \to np$ and D creation
through $ \gamma \, {}^4\text{He} \to \text{DD}$.  These are balanced
in the channel of Fig.~\ref{fig:bbn} between the ``low D'' and ``high
D'' regions, and the requirement that the electromagnetic energy that
destroys $^7$Li not disturb the D abundance specifies the preferred
decay time $\tau \sim 3\times 10^6~\s$.

Without theoretical guidance, this scenario for resolving the $^7$Li
abundance is rather fine-tuned: possible decay times and energy
releases span tens of orders of magnitude, and there is no motivation
for the specific range of parameters required to resolve BBN
discrepancies.  In the superWIMP scenario, however, both $\tau$ and
$\zetaEM$ are specified: the decay time is necessarily that of a
gravitational decay of a weak-scale mass particle, leading to
\eqref{year}, and the energy release is determined by the requirement
that superWIMPs be the dark matter, leading to \eqref{zeta}.
Remarkably, these values coincide with the best fit values for $\tau$
and $\zetaEM$.  More quantitatively, we note that the grids of
predictions for the $\Bino$ and $\stau$ scenarios given in
Fig.~\ref{fig:bbn} cover the best fit region.  Current discrepancies
in BBN light element abundances may therefore be naturally explained
by superWIMP dark matter.

This tentative evidence may be reinforced or disfavored in a number of
ways. Improvements in the BBN observations discussed above may show if
the $^7$Li abundance is truly below predictions.  In addition,
measurements of $^6$Li/H and $^6$Li/$^7$Li may constrain astrophysical
depletion of $^7$Li and may also provide additional evidence for late
decaying particles in the best fit region.\cite{Holtmann:1998gd,%
Jedamzik:1999di,Kawasaki:2000qr,Cyburt:2002uv} Finally, if the best
fit region is indeed realized by $\WIMP \to \SWIMP$ decays, there are
a number of other testable implications for cosmology and particle
physics.\cite{Feng:2003xh,Feng:2003uy} We discuss one of these in the
following section.

\section{The Cosmic Microwave Background}
\label{sec:mu}

The injection of electromagnetic energy may also distort the frequency
dependence of the CMB black body radiation.  For the decay times of
interest, with redshifts $z \sim 10^5 - 10^7$, the resulting photons
interact efficiently through $\gamma e^- \to \gamma e^-$, but photon
number is conserved, since double Compton scattering $\gamma e^- \to
\gamma \gamma e^-$ and thermal bremsstrahlung $e X \to e X \gamma$,
where $X$ is an ion, are inefficient.  The spectrum therefore relaxes
to statistical but not thermodynamic equilibrium, resulting in a
Bose-Einstein distribution function
\begin{equation}
f_{\gamma}(E) = \frac{1}{e^{E/(kT) + \mu} - 1} \ ,
\end{equation}
with chemical potential $\mu \ne 0$.

For the low values of baryon density currently favored, the effects of
double Compton scattering are more significant than those of thermal
bremsstrahlung.  The value of the chemical potential $\mu$ may
therefore be approximated for small energy releases by the analytic
expression\cite{Hu:gc}
\begin{equation}
\mu = 8.0 \times 10^{-4} 
\left[ \frac{\tau}{10^6~\s} \right]^{\frac{1}{2}} 
\left[ \frac{\zetaEM}{10^{-9}~\gev} \right] 
e^{-(\tau_{\text{dC}}/\tau)^{5/4}} \ ,
\end{equation}
where
\begin{equation}
\tau_{\text{dC}} = 6.1 \times 10^6~\s
\left[ \frac{T_0}{2.725~\K} \right] ^{-\frac{12}{5}} 
\left[ \frac{\Omega_B h^2}{0.022} \right]^{\frac{4}{5}}
\left[ \frac{1-\frac{1}{2} Y_p}{0.88} \right]^{\frac{4}{5}} \ .
\end{equation}

\begin{figure}[tbp]
\postscript{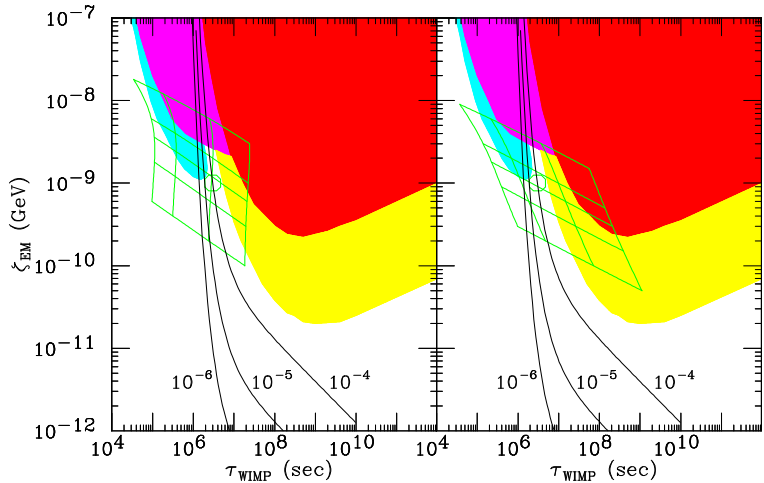}{0.85}
\caption{Contours of $\mu$, parameterizing the distortion of the CMB
  from a Planckian spectrum, in the $(\tau, \zetaEM)$ plane.  Regions
  predicted by the superWIMP dark matter scenario, and BBN excluded
  and best fit regions are given as in Fig.~\protect\ref{fig:bbn}.
\label{fig:mu} }
\end{figure}

In Fig.~\ref{fig:mu} we show contours of chemical potential $\mu$.
The current bound is $\mu < 9\times
10^{-5}$.\cite{Fixsen:1996nj,Hagiwara:fs} We see that, although there
are at present no indications of deviations from black body, current
limits are already sensitive to the superWIMP scenario, and
particularly to regions favored by the BBN considerations described in
Sec.~\ref{sec:bbn}. In the future, the Diffuse Microwave Emission
Survey (DIMES) may improve sensitivities to $\mu \approx 2 \times
10^{-6}$.\cite{DIMES} DIMES will therefore probe further into
superWIMP parameter space, and will effectively probe all of the
favored region where the $^7$Li underabundance is explained by decays
to superWIMPs.

\section{Summary and Future Directions}
\label{sec:conclusions}

SuperWIMP dark matter presents a qualitatively new dark matter
possibility realized in some of the most promising frameworks for new
physics.  In supergravity, superWIMP dark matter is realized simply by
assuming that the gravitino is the LSP.  When the NLSP is a
weakly-interacting superpartner, the gravitino superWIMP naturally
inherits the desired dark matter relic density.  The prime WIMP virtue
connecting weak scale physics with the observed dark matter density is
therefore preserved by superWIMP dark matter.

Because superWIMP dark matter interacts only gravitationally, searches
for its effects in standard dark matter experiments are hopeless.  At
the same time, this superweak interaction implies that WIMPs decaying
to it do so after BBN.  BBN observations and later observations, such
as of the CMB, therefore bracket the era of WIMP decays, and provide
new signals.  SuperWIMP and conventional WIMP dark matter therefore
have disjoint sets of signatures; we have explored the new
opportunities for dark matter detection presented by superWIMPs.  We
find that precision cosmology excludes some of the natural parameter
space, and future improvements in BBN baryometry and probes of CMB
$\mu$ distortions will extend this sensitivity.

We have also found that the decay times and energy releases generic in
the superWIMP scenario may naturally reduce $^7$Li abundances to the
observed levels without sacrificing the agreement between D and CMB
baryometry.  The currently observed $^7$Li underabundance therefore
provides evidence for the superWIMP hypothesis.  This scenario
predicts that more precise BBN observations will expose a truly
physical underabundance of $^7$Li.  In addition, probes of CMB $\mu$
distortions at the level of $\mu \sim 2 \times 10^{-6}$ will be
sensitive to the entire preferred region.  An absence of such effects
will exclude this explanation.

We have considered here the cases where neutralinos and sleptons decay
to gravitinos and electromagnetic energy.  In the case of sleptons,
BBN constraints on electromagnetic cascades provide the dominant
bound.  For neutralinos, however, the case is less clear.  Neutralinos
may produce hadronic energy through two-body decays $\chi \to Z
\Gravitino, h \Gravitino$, and three-body decays $\chi \to q\bar{q}
\Gravitino$.  Detailed BBN studies constraining hadronic energy
release may exclude such two-body decays, thereby limiting possible
neutralino WIMP candidates to photinos, or even exclude three-body
decays, thereby eliminating the neutralino WIMP scenario altogether.
At present, detailed BBN studies of hadronic energy release
incorporating the latest data are limited to decay times $\tau \alt
10^4~\s$.\cite{BBNhad} We strongly encourage detailed studies for
later times $\tau \sim 10^6~\s$, as these may have a great impact on
what superWIMP scenarios are viable.

Finally, the gravitino superWIMP scenario has strong implications for
the supersymmetric spectrum and collider searches for supersymmetry.
In particular, the possibility that the lightest observable
superpartner is charged, long thought to be excluded in supergravity
by the problems associated with charged dark matter, is in fact
viable.  Searches at hadron colliders for ``exotic'' long-lived heavy
charged particles\cite{Feng:1997zr} are therefore probes of
supergravity, the most conventional of supersymmetric theories.

\section*{Acknowledgments}
I am grateful to Arvind Rajaraman and Fumihiro Takayama for fruitful
collaboration in the work summarized here and to Pran Nath and the
organizers of SUGRA20 for creating a stimulating and enjoyable
conference.


\begin{thebibliography}{99}

\bibitem{Chamseddine:jx}
A.~H.~Chamseddine, R.~Arnowitt and P.~Nath,
%``Locally Supersymmetric Grand Unification,''
Phys.\ Rev.\ Lett.\  {\bf 49}, 970 (1982);
%%CITATION = PRLTA,49,970;%%
%\bibitem{Barbieri:1982eh}
R.~Barbieri, S.~Ferrara and C.~A.~Savoy,
%``Gauge Models With Spontaneously Broken Local Supersymmetry,''
Phys.\ Lett.\ B {\bf 119}, 343 (1982);
%%CITATION = PHLTA,B119,343;%%
%\bibitem{Hall:iz}
L.~J.~Hall, J.~Lykken and S.~Weinberg,
%``Supergravity As The Messenger Of Supersymmetry Breaking,''
Phys.\ Rev.\ D {\bf 27}, 2359 (1983);
%%CITATION = PHRVA,D27,2359;%%
%\bibitem{Alvarez-Gaume:1983gj}
L.~Alvarez-Gaume, J.~Polchinski and M.~B.~Wise,
%``Minimal Low-Energy Supergravity,''
Nucl.\ Phys.\ B {\bf 221}, 495 (1983).
%%CITATION = NUPHA,B221,495;%%

\bibitem{Goldberg:1983nd}
H.~Goldberg,
%``Constraint on the photino mass from cosmology,''
Phys.\ Rev.\ Lett.\  {\bf 50}, 1419 (1983);
%%CITATION = PRLTA,50,1419;%%
%\bibitem{Ellis:1983wd}
J.~Ellis, J.~S.~Hagelin, D.~V.~Nanopoulos and M.~Srednicki,
%``Search For Supersymmetry At The $\bar{p} p$ Collider,''
Phys.\ Lett.\  {\bf B127}, 233 (1983).
%%CITATION = PHLTA,B127,233;%%

\bibitem{Feng:2003xh}
J.~L.~Feng, A.~Rajaraman and F.~Takayama,
%``Superweakly-interacting massive particles,''
Phys.\ Rev.\ Lett.\  {\bf 91}, 011302 (2003)
[hep-ph/0302215].
%%CITATION = HEP-PH 0302215;%%

\bibitem{Feng:2003uy}
J.~L.~Feng, A.~Rajaraman and F.~Takayama,
%``SuperWIMP dark matter signals from the early universe,''
hep-ph/0306024.
%%CITATION = HEP-PH 0306024;%%

\bibitem{Krauss:1983ik}
L.~M.~Krauss,
%``New Constraints On 'Ino' Masses From Cosmology. 1. 
%Supersymmetric 'Inos',''
Nucl.\ Phys.\ B {\bf 227}, 556 (1983);
%%CITATION = NUPHA,B227,556;%%
%\bibitem{Nanopoulos:1983up}
D.~V.~Nanopoulos, K.~A.~Olive and M.~Srednicki,
%``After Primordial Inflation,''
Phys.\ Lett.\ B {\bf 127}, 30 (1983);
%%CITATION = PHLTA,B127,30;%%
%\bibitem{Khlopov:pf}
M.~Y.~Khlopov and A.~D.~Linde,
%``Is It Easy To Save The Gravitino?,''
Phys.\ Lett.\ B {\bf 138} (1984) 265;
%%CITATION = PHLTA,B138,265;%%
%\bibitem{Ellis:1984eq}
J.~R.~Ellis, J.~E.~Kim and D.~V.~Nanopoulos,
%``Cosmological Gravitino Regeneration And Decay,''
Phys.\ Lett.\ B {\bf 145}, 181 (1984);
%%CITATION = PHLTA,B145,181;%%
%\bibitem{Juszkiewicz:gg}
R.~Juszkiewicz, J.~Silk and A.~Stebbins,
%``Constraints On Cosmologically Regenerated Gravitinos,''
Phys.\ Lett.\ B {\bf 158}, 463 (1985);
%%CITATION = PHLTA,B158,463;%%
%\bibitem{Moroi:1993mb}
T.~Moroi, H.~Murayama and M.~Yamaguchi,
%``Cosmological constraints on the light stable gravitino,''
Phys.\ Lett.\ B {\bf 303}, 289 (1993);
%%CITATION = PHLTA,B303,289;%%
%\bibitem{Bolz:1998ek}
M.~Bolz, W.~Buchmuller and M.~Plumacher,
%``Baryon asymmetry and dark matter,''
Phys.\ Lett.\ B {\bf 443}, 209 (1998)
[hep-ph/9809381];
%%CITATION = HEP-PH 9809381;%%
%\bibitem{Bolz:2000fu}
%M.~Bolz, A.~Brandenburg and W.~Buchmuller,
%``Thermal production of gravitinos,''
Nucl.\ Phys.\ B {\bf 606}, 518 (2001)
[hep-ph/0012052].
%%CITATION = HEP-PH 0012052;%%

\bibitem{Spergel:2003cb}
D.~N.~Spergel {\it et al.},
%``First Year Wilkinson Microwave Anisotropy Probe (WMAP)
%Observations: Determination of Cosmological Parameters,''
astro-ph/0302209.
%%CITATION = ASTRO-PH 0302209;%%

\bibitem{Feng:2003nr}
J.~L.~Feng, A.~Rajaraman and F.~Takayama,
%``Graviton cosmology in universal extra dimensions,''
hep-ph/0307375.
%%CITATION = HEP-PH 0307375;%%

\bibitem{KKDM}
%\bibitem{Servant:2002aq}
G.~Servant and T.~M.~Tait,
%``Is the lightest Kaluza-Klein particle a viable dark matter
%candidate?,''
Nucl.\ Phys.\ B {\bf 650}, 391 (2003)
[hep-ph/0206071];
%%CITATION = HEP-PH 0206071;%%
%\bibitem{Cheng:2002ej}
H.~C.~Cheng, J.~L.~Feng and K.~T.~Matchev,
%``Kaluza-Klein dark matter,''
Phys.\ Rev.\ Lett.\  {\bf 89}, 211301 (2002)
[hep-ph/0207125];
%%CITATION = HEP-PH 0207125;%%
%\bibitem{Servant:2002hb}
G.~Servant and T.~M.~Tait,
%``Elastic scattering and direct detection of Kaluza-Klein dark
%matter,''
New J.\ Phys.\ {\bf 4}, 99 (2002)
[hep-ph/0209262];
%%CITATION = HEP-PH 0209262;%%
%\bibitem{Hooper:2002gs}
D.~Hooper and G.~D.~Kribs,
%``Probing Kaluza-Klein dark matter with neutrino telescopes,''
Phys.\ Rev.\ D {\bf 67}, 055003 (2003)
[hep-ph/0208261];
%%CITATION = HEP-PH 0208261;%%
%\bibitem{Majumdar:2002mw}
D.~Majumdar,
%``Detection rates for Kaluza-Klein dark matter,''
Phys.\ Rev.\ D {\bf 67}, 095010 (2003)
[hep-ph/0209277];
%%CITATION = HEP-PH 0209277;%%
%\bibitem{Mohapatra:2002ug}
R.~N.~Mohapatra and A.~Perez-Lorenzana,
%``Neutrino mass, proton decay and dark matter in TeV scale universal 
%extra dimension models,''
Phys.\ Rev.\ D {\bf 67}, 075015 (2003)
[hep-ph/0212254].
%%CITATION = HEP-PH 0212254;%%

\bibitem{Pagels:ke}
H.~Pagels and J.~R.~Primack,
%``Supersymmetry, Cosmology And New Tev Physics,''
Phys.\ Rev.\ Lett.\  {\bf 48}, 223 (1982).
%%CITATION = PRLTA,48,223;%%

\bibitem{Ellis:1984er}
J.~R.~Ellis, D.~V.~Nanopoulos and S.~Sarkar,
%``The Cosmology Of Decaying Gravitinos,''
Nucl.\ Phys.\ B {\bf 259}, 175 (1985).
%%CITATION = NUPHA,B259,175;%%

\bibitem{Ellis:1990nb}
J.~R.~Ellis, G.~B.~Gelmini, J.~L.~Lopez, D.~V.~Nanopoulos and S.~Sarkar,
%``Astrophysical Constraints On Massive Unstable Neutral Relic Particles,''
Nucl.\ Phys.\ B {\bf 373}, 399 (1992).
%%CITATION = NUPHA,B373,399;%%

\bibitem{Kawasaki:1994sc}
M.~Kawasaki and T.~Moroi,
%``Electromagnetic cascade in the early universe and its application 
%to the big bang nucleosynthesis,''
Astrophys.\ J.\  {\bf 452}, 506 (1995)
[astro-ph/9412055].
%%CITATION = ASTRO-PH 9412055;%%

\bibitem{Holtmann:1998gd}
E.~Holtmann, M.~Kawasaki, K.~Kohri and T.~Moroi,
%``Radiative decay of a long-lived particle and big-bang 
%nucleosynthesis,''
Phys.\ Rev.\ D {\bf 60}, 023506 (1999)
[hep-ph/9805405].
%%CITATION = HEP-PH 9805405;%%

\bibitem{Kawasaki:2000qr}
M.~Kawasaki, K.~Kohri and T.~Moroi,
%``Radiative decay of a massive particle and the non-thermal process 
%in primordial nucleosynthesis,''
Phys.\ Rev.\ D {\bf 63}, 103502 (2001)
[hep-ph/0012279].
%%CITATION = HEP-PH 0012279;%%

\bibitem{Asaka:1998ju}
T.~Asaka, J.~Hashiba, M.~Kawasaki and T.~Yanagida,
%``Spectrum of background X-rays from moduli dark matter,''
Phys.\ Rev.\ D {\bf 58}, 023507 (1998)
[hep-ph/9802271].
%%CITATION = HEP-PH 9802271;%%

\bibitem{Cyburt:2002uv}
R.~H.~Cyburt, J.~Ellis, B.~D.~Fields and K.~A.~Olive,
%``Updated nucleosynthesis constraints on unstable relic particles,''
Phys.\ Rev.\ D {\bf 67}, 103521 (2003)
[astro-ph/0211258].
%%CITATION = ASTRO-PH 0211258;%%

\bibitem{BBNhad}
%\bibitem{Reno:1987qw}
M.~H.~Reno and D.~Seckel,
%``Primordial Nucleosynthesis: The Effects Of Injecting Hadrons,''
Phys.\ Rev.\ D {\bf 37}, 3441 (1988);
%%CITATION = PHRVA,D37,3441;%%
%\bibitem{Dimopoulos:1988ue}
S.~Dimopoulos, R.~Esmailzadeh, L.~J.~Hall and G.~D.~Starkman,
%``Limits On Late Decaying Particles From Nucleosynthesis,''
Nucl.\ Phys.\ B {\bf 311}, 699 (1989);
%%CITATION = NUPHA,B311,699;%%
%\bibitem{Kohri:2001jx}
K.~Kohri,
%``Primordial nucleosynthesis and hadronic decay of a massive particle
%with a relatively short lifetime,''
Phys.\ Rev.\ D {\bf 64}, 043515 (2001)
[astro-ph/0103411].
%%CITATION = ASTRO-PH 0103411;%%

\bibitem{Hagiwara:fs}
K.~Hagiwara {\it et al.}  [Particle Data Group Collaboration],
%``Review Of Particle Physics,''
Phys.\ Rev.\ D {\bf 66}, 010001 (2002).
%%CITATION = PHRVA,D66,010001;%%

\bibitem{Kirkman:2003uv}
D.~Kirkman, D.~Tytler, N.~Suzuki, J.~M.~O'Meara and D.~Lubin,
%``The cosmological baryon density from the deuterium to hydrogen 
%ratio towards QSO absorption systems: D/H towards Q1243+3047,''
astro-ph/0302006.
%%CITATION = ASTRO-PH 0302006;%%

\bibitem{Burles:2000zk}
S.~Burles, K.~M.~Nollett and M.~S.~Turner,
%``Big-Bang Nucleosynthesis Predictions for Precision Cosmology,''
Astrophys.\ J.\  {\bf 552}, L1 (2001)
[astro-ph/0010171].
%%CITATION = ASTRO-PH 0010171;%%

\bibitem{Thorburn}
J.~A.~Thorburn,
%``The primordial lithium abundance from extreme subdwarfs: New
%observations,''
Astrophys.\ J.\  {\bf 421}, 318 (1994).

\bibitem{Bonafacio}
P.~Bonifacio and P.~Molaro,
%``The primordial lithium abundance,''
MNRAS, {\bf 285}, 847 (1997).

%\cite{Ryan:1999vr}
\bibitem{Ryan:1999vr}
S.~G.~Ryan, T.~C.~Beers, K.~A.~Olive, B.~D.~Fields and J.~E.~Norris,
%``Primordial Lithium and Big Bang Nucleosynthesis,''
Astrophys.\ J.\ Lett. {\bf 530}, L57 (2000)
[astro-ph/9905211].
%%CITATION = ASTRO-PH 9905211;%%

\bibitem{Pinsonneault:1998nf}
M.~H.~Pinsonneault, T.~P.~Walker, G.~Steigman and V.~K.~Narayanan,
%``Halo Star Lithium Depletion,''
Astrophys.\ J. {\bf 527}, 180 (1999)
[astro-ph/9803073].
%%CITATION = ASTRO-PH 9803073;%%

\bibitem{Vauclair:1998it}
S.~Vauclair and C.~Charbonnel,
%``Element Segregation in Low Metallicity Stars and the Primordial 
%Lithium Abundance,''
Astrophys.\ J. {\bf 502}, 372 (1998)
[astro-ph/9802315].
%%CITATION = ASTRO-PH 9802315;%%

\bibitem{Izotov}
Y.~I.~Izotov and T.~X.~Thuan,
%``The Primordial Abundance of 4He Revisited,''
Astrophys.\ J. {\bf 500}, 188 (1998).

\bibitem{Jedamzik:1999di}
K.~Jedamzik,
%``Lithium-6: A Probe of the Early Universe,''
Phys.\ Rev.\ Lett.\  {\bf 84}, 3248 (2000)
[astro-ph/9909445].
%%CITATION = ASTRO-PH 9909445;%%

\bibitem{Hu:gc}
W.~Hu and J.~Silk,
%``Thermalization Constraints And Spectral Distortions For 
% Massive Unstable Relic Particles,''
Phys.\ Rev.\ Lett.\  {\bf 70}, 2661 (1993).
%%CITATION = PRLTA,70,2661;%%

\bibitem{Fixsen:1996nj}
D.~J.~Fixsen \etal,
%E.~S.~Cheng, J.~M.~Gales, J.~C.~Mather, R.~A.~Shafer and E.~L.~Wright,
%``The Cosmic Microwave Background Spectrum from the Full COBE/FIRAS 
% Data Set,''
Astrophys.\ J.\  {\bf 473}, 576 (1996)
[astro-ph/9605054].
%%CITATION = ASTRO-PH 9605054;%%

\bibitem{DIMES}
http://map.gsfc.nasa.gov/DIMES.

\bibitem{Feng:1997zr}
J.~L.~Feng and T.~Moroi,
%``Tevatron signatures of long-lived charged sleptons in
%gauge-mediated supersymmetry breaking models,''
Phys.\ Rev.\ D {\bf 58}, 035001 (1998)
[hep-ph/9712499].
%%CITATION = HEP-PH 9712499;%%

\end{thebibliography}
\end{document}